\documentclass[reprint,prd,twocolumn,superscriptaddress,showpacs,showkeys,floatfix,preprintnumbers]{revtex4-1}

\usepackage{amssymb}
\usepackage{amsfonts}
\usepackage{braket}
\usepackage{xargs}
\usepackage{mathtools}
\usepackage[english]{babel}
\usepackage[utf8]{inputenc}
\usepackage{amsmath,amssymb,braket,slashed,mathrsfs}
\usepackage{pifont}
\usepackage{MnSymbol}
\usepackage{graphicx}
\graphicspath{{pic/}}
\usepackage[compat=1.1.0]{tikz-feynhand}
\usetikzlibrary{shapes,arrows,positioning}
\tikzset{decision/.style={diamond, draw, fill=blue!20, text width=4.5em, text badly centered, inner sep=0pt}}
\tikzset{block/.style={rectangle, draw, fill=blue!20, text width=10em, text centered, rounded corners, minimum width=3.5cm}}
\tikzset{block1/.style={rectangle, draw, fill=blue!20, text width=18.5em, text centered, rounded corners, minimum width=3.5cm}}
\tikzset{line/.style={draw, -latex, thick}}
\usepackage{color,hyperref}
\hypersetup{hidelinks}
\usepackage{multirow,array}
\usepackage{textcomp}

\usepackage{cleveref}
\crefname{subsection}{subsection}{subsections} 
\Crefname{subsection}{Subsection}{Subsections} 

\usepackage{changes} 

\usepackage{cancel}
\usepackage{siunitx}   
\usepackage{booktabs}   
\usepackage{subfigure}
\usepackage{balance}
\usepackage{dblfloatfix}

\newcommand{\ci}{\mathrm{i}}     
\newcommand{\ce}{\mathrm{e}}     
\newcommand{\dif}{\mathrm{d}}   

\newcommand{\abs}[1]{\left| #1 \right|} 
\newcommand{\Rmnum}[1]{\uppercase\expandafter{\romannumeral #1}} 

\newcommand{\spinup}{\uparrow}
\newcommand{\spindown}{\downarrow}

\date{\today}
\newcommand{\innovation}{Collaborative Innovation Center of Quantum Matter, Beijing 100871, China}
\newcommand{\chep}{Center for High Energy Physics, Peking University, Beijing 100871, China}
\newcommand{\pkuphy}{School of Physics, Peking University, Beijing 100871,
China}
\newcommand{\Uconn}{Department of Physics, University of Connecticut, Storrs, CT 06269, USA}

\newcommand{\SCNT}{Southern Center for Nuclear-Science Theory (SCNT), Institute of Modern Physics, Chinese Academy of Sciences, Guangdong 516000, China}

\begin{document}
\title{First-Principles Determination of the Proton-Proton Fusion Matrix Element from Lattice QCD}
\author{Zi-Yu~Wang}\affiliation{\pkuphy}\affiliation{\chep}
\author{Xu~Feng}\affiliation{\pkuphy}\affiliation{\chep}\affiliation{\innovation}\affiliation{\SCNT}
\author{Bo-Hao Jian}\affiliation{\pkuphy}
\author{Lu-Chang~Jin}\affiliation{\Uconn}
\author{Chuan~Liu}\affiliation{\pkuphy}\affiliation{\chep}\affiliation{\innovation}

\begin{abstract}
Proton–proton fusion is the fundamental weak reaction initiating stellar energy production, and a first-principles determination of its matrix element remains a long-standing goal of nuclear theory.
    We present a lattice QCD calculation of the proton-proton fusion matrix element at a pion mass of $ m_\pi \approx 432 \, \mathrm{MeV} $.
    For this process, we implement Lellouch–Lüscher finite-volume corrections within a systematic $2+\mathcal{J}\to2$ framework, explicitly accounting for two-nucleon rescattering effects, to relate finite-volume matrix elements to their infinite-volume counterparts.
Excited-state contamination is suppressed using bi-local nucleon–nucleon interpolating operators, together with a variational analysis employing operators with the three lowest momenta.
This strategy enables the determination of the two-nucleon energy spectrum and scattering parameters via L\"uscher’s finite-volume formalism.
Prior to including rescattering effects in the Lellouch–Lüscher factor, we obtain a reference value $ \braket{d | \mathcal{J} | pp}/g_A = 0.984(10)$, where $g_A$ is the nucleon axial charge.
The deviation from unity indicates a small but nonvanishing contribution from two-body currents.
    Our analysis shows that rescattering effects entering the Lellouch–Lüscher factors can substantially modify the two-body contribution, while large uncertainties in the two-nucleon scattering parameters propagate strongly into the finite-volume corrections. As a result, a precise determination of the two-body low-energy constant $L_{1,A}$ remains highly challenging with current lattice inputs.
     Despite the large uncertainty, the resulting value $L_{1,A} = 6.0(7.1) \, \mathrm{fm}^3$ is compatible, at the level of naturalness and order of magnitude, with existing phenomenological extractions from experimental data.
    This work demonstrates both the feasibility and the intrinsic challenges of ab initio lattice QCD calculations of weak two-nucleon reactions, and establishes a foundation for future studies at or near the physical pion mass.
\end{abstract}

\maketitle
\pdfbookmark{Bookmarks}{internal_label}

\section{Introduction}

The fusion of two protons into deuterium ($pp \to d e^+ \nu_e$) lies at the heart of stellar energy generation and serves as a key probe of weak interactions in low-energy nuclear environments.
As the initiating step of the proton-proton chain, this reaction governs the energy output of main-sequence stars like the Sun and, owing to its extremely slow rate, sets the timescale for stellar evolution and longevity.
From a nuclear physics perspective, proton-proton fusion offers a unique opportunity to explore the interplay between strong and weak forces, yielding fundamental insights into nuclear reaction dynamics
under astrophysical conditions.

Precise calculations of the proton–proton fusion cross-section at low energies are essential for the standard solar model.
These inputs directly affect predictions of stellar core energy release rates and
numerical simulations of neutrino production mechanisms~\cite{Bahcall:1989ks,Adelberger:2010qa}.
Its inverse process, neutrino-induced deuteron disintegration ($\bar{\nu}_e d \to nne^+$), plays a crucial role in underground neutrino detection experiments.
Improved theoretical understanding of this reaction enhances the precision of neutrino oscillation parameter extraction and supports efforts to reconstruct supernova neutrino spectra~\cite{SNO:2002tuh,Chen:2002pv,Kolbe:2003ys,Kajita:2016cak,McDonald:2016ixn}.
Moreover, it is closely related to the muon capture reaction ($\mu^- d \to nn\nu_\mu$) investigated in the MuSun experiment, which offers
stringent constraints on two-body weak interactions in effective field theories (EFTs)~\cite{MuSun:2010snd}.
Advances in theoretical modeling, lattice Quantum Chromodynamics (QCD) inputs, and precision astrophysical measurements continue to elevate
these processes as a testing ground for both fundamental physics and stellar astrophysics.

Lattice QCD has made significant progress in the computation of single-nucleon weak matrix elements.
Several groups have achieved precise determinations of the nucleon axial charge $g_A$ at physical quark masses~\cite{FlavourLatticeAveragingGroupFLAG:2024oxs},
with some attaining sub-percent level accuracy through advanced control of excited-state contamination~\cite{chang2018per}.
However, lattice QCD studies involving two-nucleon systems remain significantly more challenging.
To date, only the NPLQCD Collaboration has carried out exploratory calculations of proton-proton fusion and double beta decay at an unphysical pion mass $m_\pi\sim\SI{806}{MeV}$~\cite{Savage2017Proton-Proton,Shanahan:2017bgi,Tiburzi2017Doublebeta,Davoudi:2024ukx}.

The principal difficulty in two-nucleon studies arises from the notorious signal-to-noise problem, where correlation functions exhibit exponentially degrading signal quality with Euclidean time.
Accurate extraction of ground-state matrix elements demands enhanced overlap with the ground state, typically achieved via optimized interpolating operators.
Recent developments in two-nucleon spectroscopy include distillation~\cite{HadronSpectrum:2009krc}, stochastic Laplace-Heaviside method~\cite{Morningstar:2011ka}, sparsening field algorithms~\cite{Li_2021,Detmold:2019fbk}, and variational analysis~\cite{Francis:2018qch,amarasinghe2021variational}.
These methodologies not only improve the reliability of L\"{u}scher's finite-volume analysis but may also offer new avenues for resolving long-standing debates about the
existence of two-nucleon bound states at unphysical quark masses.

Early studies often assumed small finite-volume corrections, on the order of a few percent, based on the presence of deeply bound two-nucleon states
at heavy pion masses~\cite{Savage2017Proton-Proton,Shanahan:2017bgi,Tiburzi2017Doublebeta}.
However, for weakly bound or scattering states, finite-volume effects can be substantial.
The formalism for $2+\mathcal{J} \rightarrow 2$ processes~\cite{Briceno:2015tza,Baroni_2019,Briceno:2020vgp,Davoudi:2020xdv,Davoudi:2021noh,Moscoso:2026wmz} provides the necessary tools to account for such effects,
which we apply here to proton–proton fusion using lattice data.

This work presents a new independent lattice QCD calculation of proton-proton fusion matrix elements at $m_\pi \sim \SI{432}{MeV}$, incorporating the rescattering effects in the Lellouch-L\"uscher finite-volume factors.
To mitigate excited-state contamination, we employ bi-local interpolating operators and extract the energy spectrum using both variational analysis and multi-state fits.
Algorithmic optimizations are implemented to manage the computational complexity arising from bi-local operators at both source and sink.
This study is organized into three parts: (1) theoretical and algorithmic developments for two-nucleon interpolating operators, (2) two-nucleon spectroscopy and scattering parameter extraction, and (3) analysis of fusion matrix elements with finite-volume corrections.

\section{Methodology}
\label{sec:methodology}
In this section, we outline the methodology used in this study, including the theoretical framework, algorithmic optimizations, and finite-volume analysis.

\subsection{Two-Nucleon Interpolating Operators}
\label{subsec:interpolators}
    To construct nucleon states with the correct quantum numbers on the lattice, we use the interpolating fields as
    \begin{align}
    	\label{eq:pField}
        N_\alpha(x) &=\left[P_+\epsilon_{abc}q_{a,\alpha}(u_b^TP_+C\gamma_5d_c)\right](x),
    \end{align}
    where $a,b,c$ are color indices, $\alpha$ denotes both isospin and spin component, and the projection operator is defined as $P_+ = (1+\gamma_4)/2$.
    The quark field $q$ represents the isospin doublet $(u,d)$, and the resulting nucleon field $N$ forms the isospin doublet $(p,n)$, corresponding to the proton and neutron.
    These nucleon interpolating operators employ ``non-relativistic'' projection that reduces relativistic four-component spinors to their two-component non-relativistic counterparts.
    This reduces the computational cost of correlation functions for $A$-nucleon systems by a factor of $2^A$. In the meanwhile,
    the underlying QCD calculations are fully relativistic, and the resulting correlation functions are consistent with those obtained using four-component formulations~\cite{Doi:2011gq,Doi:2012xd}.
    As a result, this approach is widely used in lattice studies of two-nucleon systems.

   While these definitions construct individual protons or neutrons using three quark fields localized at the same spatial point, physical nucleons are extended QCD-bound states of
   quarks and gluons rather than point-like particles. To enhance overlap with the ground-state nucleon and suppress contamination from excited states,
   we apply smearing techniques that transform point sources into spatially extended wave packets with nucleon-scale distributions~\cite{APE:1987ehd}.

    Using the single-nucleon interpolating operators, the general form of a two-nucleon operator can be expressed as
    \begin{equation}
        D = \sum_{x,y}f(x,y)\sum_{\alpha,\beta}  c_{\alpha,\beta} N_\alpha(x)N_\beta(y), \label{eq:dibaryonInterpolator}
    \end{equation}
    where $c_{\alpha,\beta}$ are coefficients chosen to construct operators with the desired quantum numbers, and $f(x,y)$ encodes the spatial distribution of nucleon fields.
    By employing different forms of
    $f(x,y)$, various types of two-nucleon interpolating operators can be realized.
    It is important to consider both the implementation and computational efficiency of these operators.
    The function $f(x,y)$ can be categorized into the following representative cases:
    \begin{enumerate}
        \item Local (hexa-quark) operator $f(x,y)=\mathrm{\delta}_{x,y}$:
        This corresponds to placing both nucleons at the same spatial point. Such constructions are often referred to as ``hexa-quark'' operators in the literature and are labeled as
        ``P'' (Point) in this work.
        The double spatial sum reduces to a single sum due to the Kronecker $\mathrm{\delta}$, yielding a computational complexity of $\mathcal{O}(V)$, where $V$ is the spatial volume.

        \item Separable form $f(x,y)= \phi(x)\phi(y)$:
        In this case, the summations over $x$ and $y$ are independent
        \begin{equation}
            D = \sum_{\alpha,\beta}\left( \sum_x \phi(x) N_\alpha(x) \right) \left( \sum_y \phi(y) N_\beta(y) \right),
        \end{equation}
        maintaining a complexity of $\mathcal{O}(V)$. An example is assigning opposite momenta to the two nucleons $f(x,y)= \exp({\ci} p\cdot x) \exp(-\ci p\cdot y)$.
        A special case is $f(x,y)=1$, which corresponds to spatially separated nucleons with independent zero-momentum projections.
        These are commonly referred to as ``bi-local'' operators and are denoted as ``PS‘’ (Point-Split) in this work.

        \item Non-separable form: When $f(x,y)$ is not separable, direct evaluation in coordinate space leads to a complexity of $\mathcal{O}(V^2)$.
        A practical optimization is to use the convolution theorem via Fourier transforms, reducing the cost to $\mathcal{O}(V\ln V)$.
        This approach is used, for instance, in the calculation of two-nucleon potentials via the Nambu-Bethe-Salpeter wave function in the
        HAL QCD method~\cite{IShii_2007nuclear_forces,Aoki:2009ji,Ishii2012:hadron,Aoki:2012tk,Aoki:2012bb}.
    \end{enumerate}

    Under exact isospin symmetry, where both spin and isospin exhibit $SU(2)$ symmetry, the combination $\frac{1}{2} \otimes \frac{1}{2} = 1 \oplus 0$
    leads to two possible channels for the two-nucleon system: $^3\mathrm{S}_1$ and $^1\mathrm{S}_0$.
    These correspond to the deuteron final state and the diproton initial state in proton-proton fusion, respectively, with total spin-isospin combinations $(s, I) \in \{(1,\,0),\,(0,\,1)\}$.
    Including the spatial structure, the two-nucleon interpolating operators, constructed from single-nucleon operators, are given in Table~\ref{tab:defOfDibaryon}, from which
    the coefficients $c_{\alpha,\beta}$ can be determined.

    \begin{table}[htb]
        \centering
        \begin{tabular}{c| c}
            \hline
            Interpolators & Definitions\\
            \hline
            $pp\,(^1\mathrm{S}_0,\mathrm{P})$ & $p(\spinup,x)p(\spindown,x)$  \\
            $d\,(^3\mathrm{S}_1, s_z = 0, \mathrm{P})$ & $\frac{1}{\sqrt{2}} (p(\spinup,x)n(\spindown,x) + p(\spindown,x)n(\spinup,x))$ \\
            \hline
            $pp\,(^1\mathrm{S}_0, \mathrm{PS})$ & $\frac{1}{2}\left(p(\spinup, x)p(\spindown,y) - p(\spindown,x) p(\spinup,y)\right)$  \\
            \multirow{2}{*}{$d\,(^3\mathrm{S}_1, s_z = 0, \mathrm{PS})$} &
            $\frac{1}{2\sqrt{2}}\left(
            p(\spinup, x)n(\spindown,y) + p(\spindown,x)n(\spinup,y)\right)$ \\
            &
            $\hspace{0.5cm}- \frac{1}{2\sqrt{2}}\left(n(\spindown,x)p(\spinup,y) + n(\spinup,x)p(\spindown,y)\right)$ \\
            \hline
        \end{tabular}
         \caption{Definitions of two-nucleon interpolating operators.}
         \label{tab:defOfDibaryon}
    \end{table}

    For multi-nucleon interpolating operators, efficient implementation in lattice QCD calculations requires incorporating the Pauli exclusion principle during their construction.
    By expanding the operator in terms of quark fields and applying quark exchange symmetries to combine identical terms and eliminate vanishing contributions~\cite{Detmold:2012eu},
    this approach becomes applicable not only to two-nucleon systems but also to systems with more nucleons.
    The resulting reduced interpolating operators for an $A$-nucleon system can be written as
    \begin{eqnarray}
    O_A &=& \sum_i w_i^{A} q(\xi^i_{(1,1)})q(\xi^i_{(1,2)})q(\xi^i_{(1,3)}) \cdots
    \nonumber\\
    &&\hspace{1cm} q(\xi^i_{(A,1)})q(\xi^i_{(A,2)})q(\xi^i_{(A,3)}),
    \label{eq:QuarkLevelInterpolator}
    \end{eqnarray}
    where $w_i^A$ denotes the coefficient of the $i$-th nonvanishing quark combination, and $\xi^i_{(n,k)}$ represents the collective indices of flavor, color, spin,
and spatial coordinates of the $(n,k)$-th quark
    - where $n=1,\cdots,A$ labels the nucleon field and $k=1,2,3$ the quark within that nucleon - in the $i$-th term.
    For the dibaryon case, the coefficients listed in Table~\ref{tab:defOfDibaryon} are absorbed into $w_i^A$. In the following, we denote these coefficients by $w_i^{(NN)}$.
For a single nucleon, the corresponding coefficients are denoted by $w_i^{(N)}$.

    The number of terms in $O_A$ depends on the quantum numbers of the nucleons and the spatial distribution.
    For interpolators where all quarks reside at a single spatial point (e.g. single nucleon and hexa-quark interpolators), the numbers of terms for a single nucleon, diproton, and deuteron ($s=0$) are $9$, $21$, and $32$, respectively, compared to $12,\, 144,\, 288$ terms from a naive enumeration of indices.  The corresponding optimization ratios are $3/4$, $7/48$ and $1/9$.
    For bi-local interpolating field operators, since nucleons are positioned at distinct locations, the optimization ratio per nucleon matches that of the single-nucleon operator.
When bi-local interpolators are used at both the source and sink, the overall optimization efficiency for computing two-point correlation functions reaches $(3/4)^4 \approx 31\%$.

\subsection{Variational Method}

The variational method~\cite{Fox:1981xz,Michael:1982gb,Luscher:1990ck,ALPHA_collaboration2009GEVP}
is employed as a diagnostic and optimization tool for two-nucleon interpolators.
We focus on $s$-wave two-nucleon systems in the center-of-mass frame, labeling interpolators by the momenta of the individual nucleons: interpolators with momentum ${\bf p}=(2\pi/L)\hat{e}_x$ are denoted as $100$, while those with ${\bf p}=(2\pi/L)(\hat{e}_x+\hat{e}_y)$ are labeled as $110$, and so forth.
    We restrict the operator set to momentum-projected interpolators for two key reasons:
    \begin{itemize}
        \item Momentum eigenstates form a complete basis for representing smooth spatial distributions.
        \item High-momentum contributions, which describe short-range features, are exponentially suppressed at large Euclidean time separations.
    \end{itemize}

In the variational analysis, hexaquark operators are intentionally excluded from our operator set because they inherently contain significant contributions from a broad range of momentum states.
In practice, hexaquark interpolators exhibit limited overlap with the low-lying spectrum of interest while introducing substantially larger statistical uncertainties than momentum-space operators.
This behavior is consistent with the expectation that hexaquark operators couple indiscriminately to all momentum modes and therefore provide limited optimization for the ground state.
The NPLQCD Collaboration’s hybrid approach, which combines hexaquark and momentum-projected interpolators, supports this conclusion: their results show that hexaquark interpolators predominantly isolate individual eigenstates with significantly larger statistical noise~\cite{amarasinghe2021variational}.

    Using three lowest momentum projection, we form a $3\!\times\!3$ matrix of two-point correlation functions, $C_{ij}(t)$, and solve the GEVP:
    \begin{align}
    C_{ij}(t) &= \Braket{\Omega | \tilde{O}_i(t) \tilde{O}_j^\dagger(0) | \Omega }, \\
    C(t)v_n(t,t_\mathrm{ref}) &= \lambda_n(t,t_\mathrm{ref}) C(t_\mathrm{ref})v_n(t,t_\mathrm{ref}).
    \end{align}
    Here $\lambda_n(t,t_\mathrm{ref}) = \ce^{-E_n(t-t_\mathrm{ref})}$ corresponds to the GEVP eigenvalues, and $v_n$ denotes the eigenvector of the $n$-th state.
    The eigenenergies $E_n$ can be extracted directly from the eigenvalues $\lambda_n$.
    An alternative approach employs eigenvector projections of the correlation matrix:
    \begin{equation}
        \hat{C}_{nn} (t) =  v_n^{\dagger}(t, t_\mathrm{ref}) C(t) v_n(t, t_\mathrm{ref}).
    \end{equation}
    This projection scheme equivalently defines optimized interpolating operators for the $n$-th variational eigenstate through linear combinations of momentum-projected operators:
    \begin{equation}
        \hat{O}_n(t) = \sum_i \tilde{O}_i(t) v^i_n(t, t_\mathrm{ref}),
    \label{eq:OperatorWithEigenVector}
    \end{equation}
    where $\hat{O}_n(t)$ represents the optimized interpolator, $\tilde{O}_i(t)$ the $i$-th momentum-projected interpolators, and $v^i_n(t,t_\mathrm{ref})$ the $i$-th component of the $n$-th eigenvector.
    Although direct matrix projection and optimized operator construction are mathematically equivalent, the latter formulation allows the optimized interpolators to be used explicitly in nuclear matrix element calculations.

    To quantify momentum-state contributions to eigenstates, we define the overlap factor between the $n$-th eigenstate and $i$-th momentum state following~\cite{Bulava:2016mks}:
    \begin{equation}
        Z_{n}^i \equiv \Braket{\Omega|\tilde{O}_i(t)|n} = \left(\frac{\sum_j C_{ij}(t)v_{n}^j(t,t_\mathrm{ref})}{ \ce^{-E_n t/2} \sqrt{\hat{C}_{nn}(t)}}\right)^2.
        \label{eq:overlapFactor}
    \end{equation}

In this work, we employ the variational method to determine the finite-volume energy shift, which is particularly sensitive to excited-state contamination.
Based on the resulting GEVP analysis, we find that the ground states of both the deuteron and the dineutron/diproton systems are overwhelmingly dominated by the zero-momentum component,
with overlaps exceeding 99\%, while contributions from nonzero-momentum components are consistent with zero within statistical uncertainties (see \cref{sec:results}).
Consequently, the proton–proton fusion matrix elements are computed using zero-momentum interpolators only.
With substantially improved statistical precision in future studies, the optimized interpolators defined in Eq.~(\ref{eq:OperatorWithEigenVector}) may become advantageous for high-precision extractions of matrix elements.

\subsection{Contraction Algorithm}
    For $A$-nucleon systems, the handling of Wick contractions among $3A$ quark creation-annihilation operator pairs results in complexity that scales exponentially with the nucleon number.
    This complexity is compounded by the intricate spin and color structure between nucleons, making multi-nucleon contraction algorithms substantially more challenging than mesonic systems.
    In this work, we adapt existing algorithms for multi-nucleon contractions~\cite{Doi:2012xd,Detmold:2012eu}.
    To mitigate excited-state contamination and pseudo-plateau problems, we employ bi-local interpolators in both initial and final states.
    However, the direct application of contraction algorithms designed for hexa-quark or wall sources to bi-local sources introduces computational redundancy which can be avoided by using a more efficient contraction algorithm specifically tailored to bi-local sources.

  To optimize efficiency, we introduce the quark permutation operation $\sigma_l$, which maps a quark in the sink labeled by $(n,k)$ to a quark in the source labeled by
    $(n',k')=\sigma_l(n,k)$, where $n(n')=1,2$ label the $n(n')$-th nucleon and $k(k')=1,2,3$ label the $k(k')$-th quark within that nucleon. With this definition, the two-nucleon correlator can be reformulated as a product of three-quark building blocks, each corresponding to a single nucleon:
    \begin{align}
        C(t) &= \sum_{i, j, l} c^{l}_{ij} B_1^i(\xi^j_{\sigma_l(1,1)}, \xi^j_{\sigma_l(1,2)}, \xi^j_{\sigma_l(1,3)}) \notag \\
        &\quad B_2^i(\xi^j_{\sigma_l(2,1)},\xi^j_{\sigma_l(2,2)},\xi^j_{\sigma_l(2,3)}), \label{eq:algForBilocal} \\
        c^l_{ij} &= W_i w_j^{(NN)} \operatorname{sgn}(\sigma_l).
    \end{align}
    Here, $c^l_{ij}$ encapsulates dibaryon final-state coefficients $W_i$ (listed in Table~\ref{tab:defOfDibaryon}), quark-level initial-state coefficients $w_j^{(NN)}$ (defined in Eq.~(\ref{eq:QuarkLevelInterpolator}))
    and the sign associated with the quark permutation $\sigma_l$ arising from Wick contractions.
    The quantity $B_n^i$ represents the $n$-th nucleon in the $i$-th final-state, which is built from the product of three quark propagators,
    \begin{eqnarray}
        &&B^i_n(\xi^j_{\sigma_l(n,1)}, \xi^j_{\sigma_l(n,2)}, \xi^j_{\sigma_l(n,3)})
        \nonumber\\
         &&\hspace{1cm}=\sum_{j'} w_{j'}^{(N)} \prod_{k=1}^3 S(\xi^{j'}_{(n,k)},\xi^j_{\sigma_l(n,k)}),
    \end{eqnarray}
    where $S(\xi^{j'}_{(n,k)},\xi^j_{\sigma_l(n,k)})$ denotes the quark propagator from an initial-state quark $\sigma_l(n,k)$ to a final-state quark $(n,k)$ and $w_{j'}^{(N)}$ are the quark-level final-state coefficients.
Note that the sink and source operators are treated differently. In the sink, the three-quark block structure is preserved, and the corresponding coefficients $W_i$ and $w_{j'}^{(N)}$
 are retained explicitly. In contrast, in the source all six quarks are fully expanded with coefficients $w_j^{(NN)}$, into which the dibaryon coefficients $W_i$ have been absorbed.
 This formulation extends the Unified Contraction Algorithm~\cite{Doi:2012xd} to point-splitting (bi-local) interpolating operators, explicitly accommodating spatially separated nucleons.

    We further improve efficiency by classifying quark permutations into intra-nucleon exchanges (within a single nucleon) and inter-nucleon exchanges (between different nucleons). Intra-nucleon permutations are absorbed into the three-quark blocks, leading to a significant reduction in the size of the coefficient tables. For instance, for a proton three-quark block, the up-quark permutation group is reduced from $S_{n_u}$ to $S_{n_u}/S_2$,
    , which decreases the table size by a factor of four in two-nucleon systems.
Moreover, three-quark blocks are shared among multiple contraction terms and across different dibaryon channels. By computing each distinct block only once, redundant calculations are avoided. The resulting efficiency gain is quantified by the redundancy factor $\eta$ reported in Table~\ref{tab:3quarkBlockComp}.

    \begin{table}[htbp]
        \centering
        \begin{tabular}{c|ccc}
            \hline
            Dibaryon System & $N_\mathrm{table}$ & $N_\mathrm{block}$ & $\eta \equiv N_\mathrm{block}/N_\mathrm{table}$ \\
            \hline
            $pp(^1\mathrm{S}_0)$ & 1944 & 468 & 0.24 \\
            $pn(^1\mathrm{S}_0)$ & 2916 & 468 & 0.16 \\
            $d(^3\mathrm{S}_1)$ & 2916 & 468 & 0.16 \\
            \hline
        \end{tabular}
                \caption{Efficiency gain for the contraction algorithm applied to bi-local dibaryon operators. $N_\mathrm{table}$: number of entries in the reduced coefficient table; $N_\mathrm{block}$: number of unique three-quark blocks; $\eta$: redundancy factor, with smaller values indicating higher efficiency.}
                   \label{tab:3quarkBlockComp}
    \end{table}

\subsection{Lattice Computation of Weak Decay Matrix Elements}\label{subsec:NMEinLQCD}

Weak decay matrix elements are extracted from three-point correlation functions with weak-current insertions in lattice QCD, using bi-local interpolating operators.
To directly exploit the optimized algorithms described in the previous subsection, we employ the sequential-source propagator (background-field) technique in combination with the summation method, which enhances the signal-to-noise ratio and suppresses excited-state contamination.
    The sequential source propagator technique has been broadly used in lattice studies of electromagnetic and weak currents~\cite{Fucito:1982gAquenched,Martinelli:1982moments,Bernard:1982Moments,Savage2017Proton-Proton,Tiburzi2017Doublebeta,Shanahan:2017bgi}.

    While it is in principle feasible to explicitly incorporate the Wick contractions associated with the weak current into the contraction algorithm, a simpler and equally efficient approach exploits the exact isospin
    symmetry of lattice QCD.
    According to the Wigner-Eckart theorem, matrix elements of tensor operators satisfy
    \begin{align}
        &\Bra{\alpha',j'm'}T^{(k)}_q \Ket{\alpha, jm} \notag \\
        &= \Braket{jk;mq | jk;j'm'} \frac{\Braket{\alpha', j' || T^{(k)} || \alpha, j}}{\sqrt{2j+1}},
    \end{align}
    where $\Braket{jk;mq | jk;j'm'}$ denotes the Clebsch-Gordan coefficient, and the reduced matrix element depends only on the total quantum numbers $j,j',k$, but not on their projections $m,m',q$.
    This relation allows flavor-changing weak decay matrix elements to be reconstructed from isospin-related, flavor-neutral current matrix elements using Clebsch-Gordan coefficients.
Consequently, the same contraction algorithms developed for two-point correlators can be applied directly to three-point correlation functions with weak current insertions.

    For single beta decay with an isospin change $\Delta I=1$, the nucleon-to-proton transition ($I_z=-1/2\to I_z=1/2$) can be expressed in terms of $1/2 \otimes 1$ Clebsch-Gordan coefficients,
    relating it to proton-to-proton matrix elements of isospin-neutral currents.
    In proton-proton fusion, the initial $I=1$ triplet state transitions to an $I=0$ singlet final state. This transition is described by the $1 \otimes 1$ Clebsch-Gordan coefficients
    relevant for the $pn(^1\mathrm{S}_0)\rightarrow pn(^3\mathrm{S}_1)$ process. The resulting relations between matrix elements of flavor-changing and flavor-conserving currents are given by:
    \begin{align}
        \braket{p|J^+|n} &= \sqrt{2} \braket{p|J^3|p}, \\
        \braket{d|J^+|pp(^1\mathrm{S}_0)} &= -\braket{d|J^3|pn(^1\mathrm{S}_0)}.
    \end{align}

    Assuming exact isospin symmetry, we define the flavor-conserving and flavor-changing axial currents as
        \begin{equation}
        J^a(x) =\frac{1}{2}\overline{u}(x)\Gamma\tau^a u(x) , \quad \Gamma=\gamma_3\gamma_5\label{eq:J3} \\
    \end{equation}
    with $\tau^{\pm}=\frac{1}{\sqrt{2}}(\tau_1\pm i\tau_2)$ and $\tau$ denotes
Pauli matrices that act in isospin space.
    \footnote{This convention for both neutral and charged currents is adopted by the NPLQCD collaboration in their study of double-beta decay~\cite{Tiburzi2017Doublebeta}.
    In their related work on proton–proton fusion~\cite{Savage2017Proton-Proton}, however, the neutral current $J^3$ is defined without the factor of $1/2$.}

    The sequential-source propagator technique implements these currents through modified quark propagators:
    \begin{align}
        S^u_\lambda(y,x) &= S_0 (y, x) +  \frac{\lambda}{2} \int \dif z S_0 (y, z) \Gamma S_0(z, x), \\
        S^d_\lambda(y,x) &= S_0 (y, x) -  \frac{\lambda}{2} \int \dif z S_0 (y, z) \Gamma S_0(z, x).
    \end{align}
    Here $S_0(y, x)$ denotes quark propagators without current insertions. The resulting correlators become $\lambda^6$-order polynomials for two-nucleon systems. Three-point correlators are then computed via
        \begin{align}
            R_3(t) &\equiv \frac{C_3(t)}{C_2(t)} = \frac{C_{+\lambda}(t) - C_{-\lambda}(t)}{2\lambda C_{\lambda=0}(t)}, \label{eq:correlationSeq3pt}
        \end{align}
    with $\lambda=0.01$ yielding truncation errors $\mathcal{O}(\lambda^{2})\sim 10^{-4}$, negligible compared to our $1\%-10\%$ uncertainties of matrix elements.

    Analyzing the time dependence of $R_3(t)$ defined in Eq.~(\ref{eq:correlationSeq3pt}), it yields~\cite{Bouchard_2017FHT}:
    \begin{align}
        C_3(t) &= \sum_{t^{\prime}=0}^{T-1}\Braket{ \Omega\left|T\left\{\mathcal{O}(t) \mathcal{J}\left(t^{\prime}\right) \mathcal{O}^{\dagger}(0)\right\}\right| \Omega}, \notag\\
        &= \sum_{n} \mathrm{e}^{-E_n t} \left( (t-1)J_{nn} Z^{\bf{0}}_n Z^{\dagger}_n + d_n \right) \notag \\
        &\quad + \sum_{m\neq n} Z_{n}^{\bf{0}} Z_{m}^{\dagger} \frac{\mathrm{e}^{-E_{n} t-\frac{\Delta_{m n}}{2}}-\mathrm{e}^{-E_{m} t-\frac{\Delta_{n m}}{2}}}{\mathrm{e}^{\frac{\Delta_{m n}}{2}}-\mathrm{e}^{\frac{\Delta_{n m}}{2}}} J_{n m}, \notag \\
        R_3(t) &= \frac{C_3(t)}{C_2(t)} = \left( \frac{\Braket{f|\mathcal{J}|i}}{Z_A} \right) t + \cdots, \label{eq:ratioThreePoint}
    \end{align}
    where $Z^{\bf{p}}_n = \sum_{\bf{x}}  \mathrm{e}^{\mathrm{i} {\bf p} \cdot {\bf x}} \Braket{\Omega | \mathcal{O} (0,{\bf x}) | n}$ and $Z^{\dagger}_n = \Braket{n | \mathcal{O}^\dagger (0,{\bf 0}) | \Omega}$ are the coupling factors between the interpolator and nucleon eigenstates, $J_{nm}=\Braket{n|\mathcal{J}|m}$ represents the matrix element corresponding to the current $\mathcal{J} \equiv \mathcal{J}(0) = \int \mathrm{d}^3 {\bf x}\, J(t, {\bf x})$. $\Delta_{mn} = E_m-E_n$ is the energy difference between initial and final states, and $d_n$ denotes time-independent constant terms arising from current insertions outside the source-sink time interval. The ellipsis in Eq.~(\ref{eq:ratioThreePoint}) represents time-independent contributions and exponentially suppressed terms.

\subsection{Finite-Volume Analysis}
\label{subsec:FVAnalysis}

Existing lattice calculations of proton-proton fusion~\cite{Savage2017Proton-Proton} have treated corrections from two-nucleon rescattering under the assumption that, at the heavy pion mass used in that work ($m_\pi \sim \SI{806}{MeV}$), the two-nucleon system forms deeply bound states, making such corrections relatively small. However, the existence of these bound states at heavy pion masses remains an open question. As shown in~\cref{sec:results}, our study at $m_\pi \sim \SI{432}{MeV}$ suggests that the system may instead be shallowly bound or lie in a scattering regime, where finite-volume effects could become more significant.

Finite-volume effects associated with single-current insertions in two-body processes have been extensively studied~\cite{Briceno:2015tza,Baroni_2019,Briceno:2020vgp}, and the matching between finite-volume lattice matrix elements for proton-proton fusion and the corresponding low-energy constants (LECs) in pionless effective field theory ($\cancel{\pi}\mathrm{EFT}$) has been established~\cite{Davoudi:2020xdv,Davoudi:2021noh,Moscoso:2026wmz}.
Since $\cancel{\pi}\mathrm{EFT}$ is formulated as a non-relativistic theory in infinite volume, consistent normalization of nuclear states is required for the matching. We therefore employ a non-relativistic normalization for lattice nuclear states
\begin{align}
        \Braket{m, NN|n, NN}_L  &= \delta_{m,n}, \\
        \Braket{{\bf p},NN|{\bf p}',NN}_{\infty} &= (2\pi)^3 \delta^{(3)}({\bf p} - {\bf p}'),
    \end{align}
    where ${\bf p}$ and ${\bf p}'$ denote the single-nucleon momenta in the center-of-mass frame.

    The finite-volume correction for $2+\mathcal{J}\rightarrow 2$ processes is \cite{Briceno:2015tza,Davoudi:2020xdv,Davoudi:2021noh}:
    \begin{align}
        &L^6 \abs{\Braket{d,E_f,L|J^-|pp,E_i,L}}^2 \notag \\
        &= \abs{\mathcal{R}^{(d)}(E_f)} \abs{\mathcal{W}_{\mathrm{df,V}}(E_i,E_f)}^2\abs{\mathcal{R}^{(pp)}(E_i)}, \label{eq:FV3ptAll}
    \end{align}
    where $\mathcal{W}_{\mathrm{df,V}}(E_i,E_f)$ is the finite-volume divergence-free amplitude \cite{Briceno:2015tza}, and the generalized Lellouch-L\"{u}scher factor $\mathcal{R}(E)$ is \cite{Christ:2014qaa,Christ:2015pwa}:
    \begin{align}
        \mathcal{R}(E_n)
        &= \lim_{E\rightarrow E_n} \frac{E-E_n}{F_0^{-1}(E) + \mathcal{M}(E)}, \notag\\
        &=\left[\frac{\dif}{\dif E}\left(F^{-1}_0 (E) + \mathcal{M}(E)\right)\right]_{E=E_n}^{-1}, \notag\\
        &= -\mathcal{M}^{-2}(E_n) \left[\frac{\dif}{\dif E}\left(F_0 (E) + \mathcal{M}^{-1}(E)\right)\right]_{E=E_n}^{-1},\label{eq:llfactorDef}
    \end{align}
    with scattering amplitude $\mathcal{M}(E)$ and the finite-volume function $F_0(E^+)$ related to the integral $I_0(E^+)$:
    \begin{align}
        \mathcal{M}(E) &= \frac{4\pi}{M}\frac{1}{p \cot \delta - \ci p}, \label{eq:MDef}\\
        I_0(E^+) &= \int \frac{\dif^3 \bm{k}}{(2\pi)^3} \frac{1}{E - \frac{\bm{k}^2}{M}+ \ci \epsilon}, \\
        F_0(E^+) &= \left(\frac{1}{L^3} \sum_{\bm{k}} - \int \frac{\dif^3 \bm{k}}{(2\pi)^3} \right) \frac{1}{E - \frac{\bm{k}^2}{M}+ \ci \epsilon} ,\notag \\
        &=\frac{M}{4\pi} \left(-\frac{\sqrt{4\pi}}{\pi L}\mathcal{Z}_{00}\left(1;\left(\frac{pL}{2\pi}\right)^2\right) + \ci p\right),\label{eq:F0Def}\\
        \mathcal{Z}_{00}(s;q^2) &= \frac{1}{\sqrt{4\pi}} \sum_{\bm{n}\in \mathbb{Z}^3} \frac{1}{\left(\bm{n}^2 - q^2\right)^s}.
    \end{align}
    Here $p=\sqrt{ME}$ is the relative momentum carried by each nucleon.

    The above expressions are written in the nonrelativistic limit. In the relativistic formulation, the momentum $p$ is related to the total center-of-mass energy by the dispersion relation $E=2\sqrt{M^2+p^2}$.
    The relativistic formalism can be recovered by rescaling the scattering amplitude as $\mathcal{M} \to 2EM \mathcal{M}$ and replacing the nonrelativistic propagator $\left(E-\frac{k^2}{M}+i\epsilon\right)^{-1}$
    with its relativistic counterpart $E^{-1}(2\omega_k)^{-1}\left(E-2\omega_k+i\epsilon\right)^{-1}$, where $\omega_k=\sqrt{M^2+k^2}$.
    For the ground state, where the dibaryon lies very close to threshold, the difference between the nonrelativistic and relativistic formulations is negligible. However, when combining the ground and excited states to extract the scattering length and effective range, relativistic kinematics become important. We therefore employ the relativistic formalism for the scattering analysis, while the nonrelativistic framework is used only for matching the matrix elements to $\cancel{\pi}\mathrm{EFT}$.

    For dibaryon systems, the Lellouch-L\"uscher factors must be determined directly from derivatives of the phase shift, since the scattering lengths are too large for the large-$L$ expansion to be reliable. This contrasts with mesonic systems, for which the large-$L$ expansion typically works very well.
    Expanding Eq.~(\ref{eq:llfactorDef}) by subsitituting Eqs.~(\ref{eq:MDef}) and (\ref{eq:F0Def}) follows:
    \begin{align}
        \mathcal{R}(E) &= \frac{1}{L^3}\frac{4\pi \mathcal{Z}^2_{00}(1;q^2) + 4\pi^4 q^2}{\sqrt{4\pi}\mathcal{Z}_{00}(2;q^2) - 2\pi^2 \frac{\dif}{\dif q^2} q \cot \delta(q)},\label{eq:llfactor}
    \end{align}
    where $q= pL/2\pi$ is the dimensionless momentum, the term of $q \cot \delta$ can be determined from lattice calculations by L\"{u}scher's method \cite{Luscher:1985dn,Luscher:1986pf,Luscher:1990ck,Luscher:1991cf}, and the zeta function $\mathcal{Z}_{00}(s;q^2)$ is computed numerically \cite{Luscher:1986pf,Luscher:1990ux,Feng:2004ua}.

    The relationship between finite-volume and infinite-volume amplitudes is \cite{Davoudi:2020xdv}:
    \begin{align}
        &\ci \mathcal{W}_{\mathrm{df,V}} (E_i,E_f) = \ci \mathcal{W}_{\mathrm{df}} (E_i,E_f)  \notag \\
        &+ \ci\mathcal{M}^{(d)}(E_f) \left[\ci F_A((p_f-p_i)^2) F_1(E_f,E_i)\right] \ci \mathcal{M}^{(pp)}(E_i), \label{eq:WdfVtoWdf}
    \end{align}
    where $F_A((p_f-p_i)^2)$ denotes the on-shell axial form factor associated with the weak current inserted on a single-nucleon line. At zero momentum transfer, $(p_f-p_i)^2=0$, it reduces to $F_A(0)=g_A$. The triangle-diagram functions, $F_1(E_i, E_f)$ and $I_1(E_i, E_f)$, arising from three-propagator loop integrals, are defined as
    \begin{align}
        I_1(E_i, E_f) &\equiv \int \frac{\dif^3 \bm{k}}{(2\pi)^3} \frac{1}{E_i - \frac{\bm{k}^2}{M} + \ci \epsilon} \frac{1}{E_f - \frac{\bm{k}^2}{M} + \ci \epsilon} ,\notag \\
        &= \frac{1}{E_f -E_i} \left(I_0(E_i)-I_0(E_f)\right), \\
        F_1(E_i, E_f)
        &= \frac{F_0(E_i)-F_0(E_f)}{E_f-E_i}. \label{eq:F1ToF0}
    \end{align}
    Using Eq.~(\ref{eq:WdfVtoWdf}), the infinite-volume amplitude $\mathcal{W}_{\mathrm{df}} (E_i,E_f)$ can be determined from the finite-volume result.

\subsection{Matching with EFT}

In EFT~\cite{Butler:2002cw}, the deuteron and diproton are represented by the projection operators
$P_{i}=\sigma_{2} \sigma_{i} \tau_{2} / \sqrt{8}$ and $\tilde{P}_{+}=\sigma_{2} \tau_{2} \tau_{+} / \sqrt{8}$, respectively, where $\sigma$ and $\tau$ denote the Pauli matrices acting in spin and isospin space.
Expanding these matrices explicitly shows that these operators are the same as those listed in Table~\ref{tab:defOfDibaryon}.

    The weak current in EFT is given by~\cite{Butler:2002cw}
    \begin{align}
        A_{k}^{-}= & A_{k(1)}^{-} +A_{k(2)}^{-} \notag \\
        =&\frac{g_{A}}{2} N^{\dagger} \tau_{-} \sigma_{k} N
        +L_{1, A}\left[\left(N^{T} P_{k} N\right)^{\dagger}\left(N^{T} \tilde{P}_{-} N\right)+\mathrm{h.c.}\right] \notag \\
        & +\cdots, \label{eq:AkInEFT}
    \end{align}
    where $L_{1,A}$ is a two-body low-energy constant (LEC), and the ellipsis denotes higher-order operators.
   Both operators convert a spin-triplet state into a spin-singlet state, leading to the matrix elements
    \begin{eqnarray}
        \Braket{{\bf q}_f, d|A_{k(1)}^{-} \otimes {\bf 1}| {\bf q}_i, pp} &= &\frac{1}{2} g_A \delta^{3}({\bf q}_f - {\bf q}_i), \label{eq:AkgA}\\
        \Braket{{\bf q}_f, d|A_{k(2)}^{-}| {\bf q}_i, pp} &=& L_{1,A} \label{eq:AkL1A},
    \end{eqnarray}
    where ${\bf q}_i$ denotes the momentum of one nucleon in the center-of-mass frame (equivalently, the relative momentum of the $pp$ system), and ${\bf q}_f$ denotes the momentum carried by the same nucleon in the final state.

    \begin{figure}[htbp]
    \centering
    \subfigure[c][One-body]{ %
        \label{spic:gA}
        \begin{tikzpicture}
            \setlength{\feynhandlinesize}{1pt}
            \begin{feynhand}
                \vertex [NWblob] (a) at (0,0) {};
                \vertex [NWblob] (b) at (4,0) {};
                \vertex [crossdot] (c) at (2,1) {};
                \vertex (d) at (2,2) {};
                \propag[plain] (a) to [in=180, out=80, looseness=0.7] (c);
                \propag[plain] (c) to [in=100, out=0, looseness=0.7] (b);
                \propag[plain] (b) to [in=300, out=240, looseness=0.4] (a);
                \propag[bos] (c) to (d);
            \end{feynhand}
        \end{tikzpicture}
    }
    \subfigure[c][Two-body]{ 
        \label{spic:l1A}
        \begin{tikzpicture}
            \setlength{\feynhandlinesize}{1pt}
            \begin{feynhand}
                \vertex [NWblob] (a) at (0,0) {};
                \vertex [NWblob] (b) at (4,0) {};
                \vertex [crossdot] (c) at (2,0) {};
                \vertex (d) at (2,1) {};
                \propag[plain] (a) to [in=120, out=60, looseness=0.75] (c);
                \propag[plain] (c) to [in=300, out=240, looseness=1] (a);
                \propag[plain] (b) to [in=60, out=120, looseness=0.75] (c);
                \propag[plain] (c) to [in=240, out=300, looseness=1] (b);
                \propag[bos] (c) to (d);
            \end{feynhand}
        \end{tikzpicture}
    }
    \caption{One-body contribution$(a)$ and two-body contribution $(b)$.}\label{pic:ppFusionInEFT}
    \end{figure}
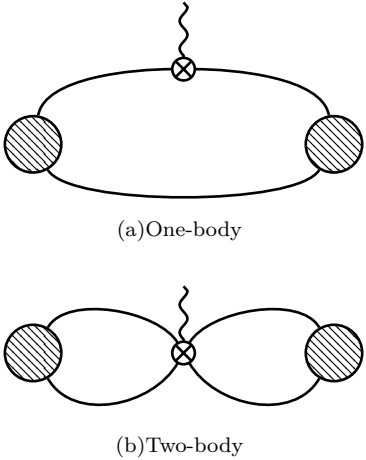

        The infinite-volume, divergence-free amplitude can be decomposed into single-body and two-body contributions (see Fig.~\ref{pic:ppFusionInEFT}):
    \begin{align}
        \ci \mathcal{W}_{\mathrm{df}}(E_i,E_f) &= \ci\mathcal{W}_{\mathrm{df}}^{\mathrm{1B}}(E_i,E_f) + \ci\mathcal{W}^{\mathrm{2B}}(E_i,E_f), \label{eq:Wdf} \\
        \ci\mathcal{W}_{\mathrm{df}}^{\mathrm{1B}}(E_i,E_f) &= \ci \mathcal{M}(E_f) \left[\ci g_A I_1(E_i,E_f)\right] \ci \mathcal{M}(E_i), \label{eq:W1B}\\
        \ci\mathcal{W}^{\mathrm{2B}}(E_i,E_f) &= \ci \mathcal{M}(E_f) \left[\ci \tilde{L}_{1,A}\right] \ci \mathcal{M}(E_i), \label{eq:W2B}
    \end{align}
    where $I_1$ arises from three-propagator loop integrals, and $\tilde{L}_{1,A}$ is a scale-independent LEC defined by
    \begin{align}
        \tilde{L}_{1,A} &= \frac{1}{C^{(d)}_0 C^{(pp)}_0} \left[L_{1,A}- \frac{g_A M}{2}\left(C^{(d)}_2 +C^{(pp)}_2\right)\right].\label{eq:L1ARelation}
    \end{align}
    Here $C^{(d)}_0, C^{(d)}_2, C^{(pp)}_0, C^{(pp)}_2$ are the LECs in $\cancel{\pi}\mathrm{EFT}$~\cite{Davoudi:2020xdv,Davoudi:2021noh}:
    \begin{align}
        C^{(d)}_{0}(\mu) &=\frac{4 \pi}{M} \frac{1}{(-\mu+1 / a^{(d)})}, \\
        C^{(d)}_{2}(\mu) &=\frac{2 \pi}{M} \frac{r^{(d)}}{(-\mu+1 / a^{(d)})^2} \\
        C^{(pp)}_{0}(\mu) &=\frac{4 \pi}{M} \frac{1}{(-\mu+1 / a^{(pp)})}, \\
        C^{(pp)}_{2}(\mu) &=\frac{2 \pi}{M} \frac{r^{(pp)}}{(-\mu+1 / a^{(pp)})^2},
    \end{align}
    where $\mu$ is the renormalization scale, $a$ is the scattering length and $r$ is the effective range.

    Using this formalism together with numerical results from lattice QCD calculations, one can first determine the scale-independent low-energy constant
    $\tilde{L}_{1,A}$, and subsequently constrain $L_{1,A}$ via Eq.~(\ref{eq:L1ARelation}).

    Note that Eqs.~(\ref{eq:Wdf}) - (\ref{eq:W2B}) do not hold in general, since the divergence-free amplitude $\mathcal{W}_{\mathrm{df}}(E_i,E_f)$ cannot be factorized into a product of on-shell scattering amplitudes of the form $\mathcal{M}(E_f) [\cdot] \mathcal{M}(E_i)$.
    These relations are valid only within the framework of next-to-leading-order (NLO) $\cancel{\pi}\mathrm{EFT}$~\cite{Davoudi:2020xdv}.
    Following this approach, and using Eqs.~(\ref{eq:FV3ptAll}), (\ref{eq:WdfVtoWdf}), and (\ref{eq:W1B}), we obtain the one-body contribution
    \begin{align}
        &L^6\abs{\Braket{E_f,L|\mathcal{J}|E_i,L}}^2_{\mathrm{1B}} \notag \\
        &\approx g_A^2 \frac{\left[\left(\sqrt{4\pi}\mathcal{Z}_{00}(1;q_i^2)-\sqrt{4\pi}\mathcal{Z}_{00}(1;q_f^2)\right)/\left(q_i^2-q^2_f\right)\right]^2}{\left(\sqrt{4\pi}\mathcal{Z}_{00}(2;q_i^2) -\pi^2 R_i\right)\left(\sqrt{4\pi}\mathcal{Z}_{00}(2;q_f^2) -\pi^2 R_f\right)},\label{eq:FV1B}
    \end{align}
    where $ R = 2\pi r_0/L$ is the dimensionless effective range, determined from the effective-range expansion (ERE)
    \begin{equation}
        p \cot \delta(p) = -\frac{1}{a} +\frac{1}{2} r_0 p^2.
        \label{eq:ERE}
    \end{equation}
    with $a$ denoting the scattering length and $r_0$ the effective range.

    For the two-body contribution, Eqs.~(\ref{eq:FV3ptAll}), (\ref{eq:WdfVtoWdf}), and (\ref{eq:W2B}) lead to
    \begin{align}
        &L^6 \abs{\Braket{d,E_f,L|\mathcal{J}|pp,E_i,L}}^2_{\mathrm{2B}} \notag \\
        &= \abs{\mathcal{R}^{(d)}(E_f)} \abs{\mathcal{W}_{\mathrm{2B}}(E_i,E_f)}^2\abs{\mathcal{R}^{(pp)}(E_i)}. \label{eq:FV2B}
    \end{align}

\section{Numerical Results}\label{sec:results}
    \subsection{Lattice Parameters}
    We utilize $2+1$-flavor domain-wall fermion ensembles generated by the RBC/UKQCD collaboration with Iwasaki gauge action. The lattice volume is $L^3 \times T = 24^3 \times 64$ with lattice spacing $a \sim \SI{0.11}{fm}$, corresponding to a spatial extent of $\SI{2.65}{fm}$. The pion mass is $m_\pi = 432.2(1.4)\,\mathrm{MeV}$ and axial vector current renormalization factor $Z_A=0.71759(56)$~\cite{Aoki_2011_Continuum_limit,Boyle_2016_LECs,Yamazaki_2008,RBC:2007yjf}.

    We analyze 162 gauge configurations and adopt the sparsening-fields method~\cite{Li_2021,Detmold:2019fbk} by placing eight smeared sources on each even time slice $t=0,2,4,\cdots,62$.
    For each configuration, light quark propagators are computed from all eight smeared sources.
Sequential-source propagators implementing the current insertion are then constructed using four of these smeared-source propagators on each time slice.

    \begin{table}[htbp]
        \centering
        \begin{tabular}{c c c c c}
            \hline
            $L^3\times T$ & $a^{-1}(\mathrm{GeV})$ & $m_{\pi}(\mathrm{MeV})$ & meas. & meas.(seq.)\\
            \hline
            $24^3 \times 64$ & $1.7844(49)$ & $432.2(1.4)$ & $162 \times 256$ & $162 \times 128$  \\
            \hline
        \end{tabular}
        \caption{Parameters of lattice gauge ensemble used in this work.}
        \label{tab:latticeSetup34pt}
    \end{table}

    \subsection{Two-Nucleon Spectrum}
    \label{subsec:spectrum}

    \begin{figure}[htbp]
        \centering
        \includegraphics[width = \linewidth]{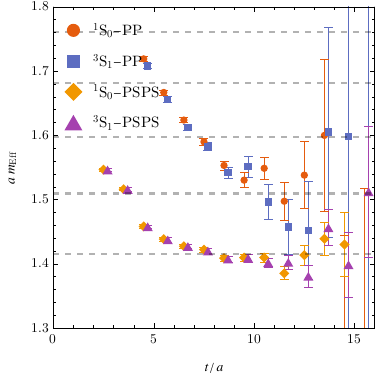}
        \caption{Effective mass curves obtained using different dinucleon interpolators. PP denotes hexaquark interpolators at both the source and sink, while PSPS corresponds to bi-local interpolators at both source and sink. Data points for the $^1\mathrm{S}_0$ and $^3\mathrm{S}_1$ channels are slightly offset horizontally for clarity.}
        \label{pic:2ptPPvsPSPS}
    \end{figure}

    We first compare the effects of different interpolators on the dinucleon energy spectrum, then perform L\"{u}scher finite-volume analysis using current data as input.
    The effective mass curves obtained using hexa-quark and bi-local interpolators are shown in Fig.~\ref{pic:2ptPPvsPSPS}.
    While correlators constructed from both interpolators converge at large Euclidean times, those built from hexa-quark interpolators exhibit significantly larger excited-state contamination at early times.
    This behavior originates from the enforced spatial overlap $f(x,y) = \delta_{x,y}$, which assigns equal weight to all momentum modes and thereby enhances excited-state contributions. In addition to causing the familiar
    ``pseudo plateau'' problem in effective-mass determinations, previous lattice QCD studies of single-nucleon weak decays have shown that such contamination can also bias matrix-element extractions based on ratios of three-point to two-point functions~\cite{FlavourLatticeAveragingGroupFLAG:2024oxs}. These considerations motivate our use of computationally more demanding bi-local interpolators to suppress excited-state effects.

    \begin{figure}[htbp]
        \centering
        \includegraphics[width = \linewidth]{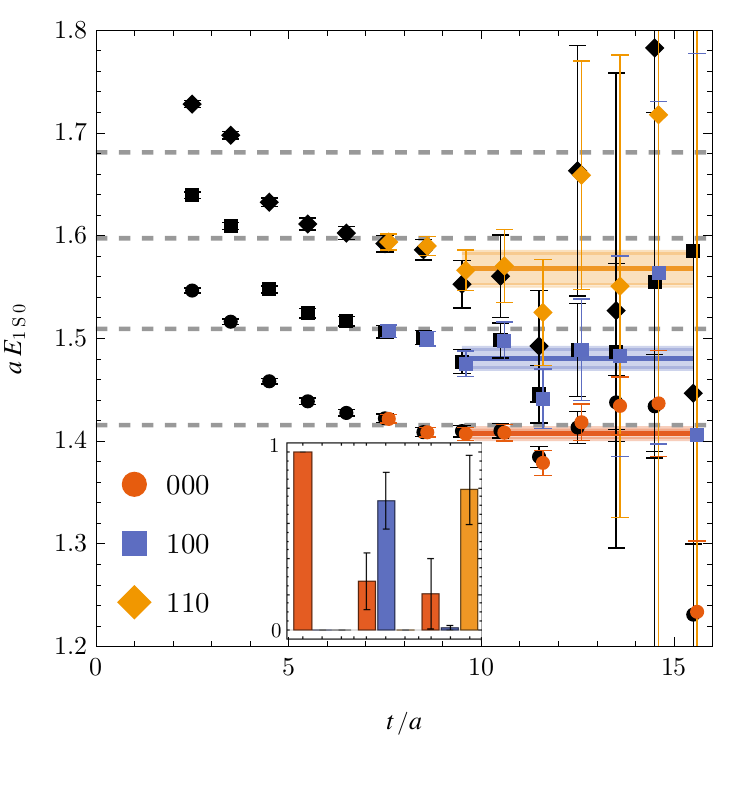}
        \includegraphics[width = \linewidth]{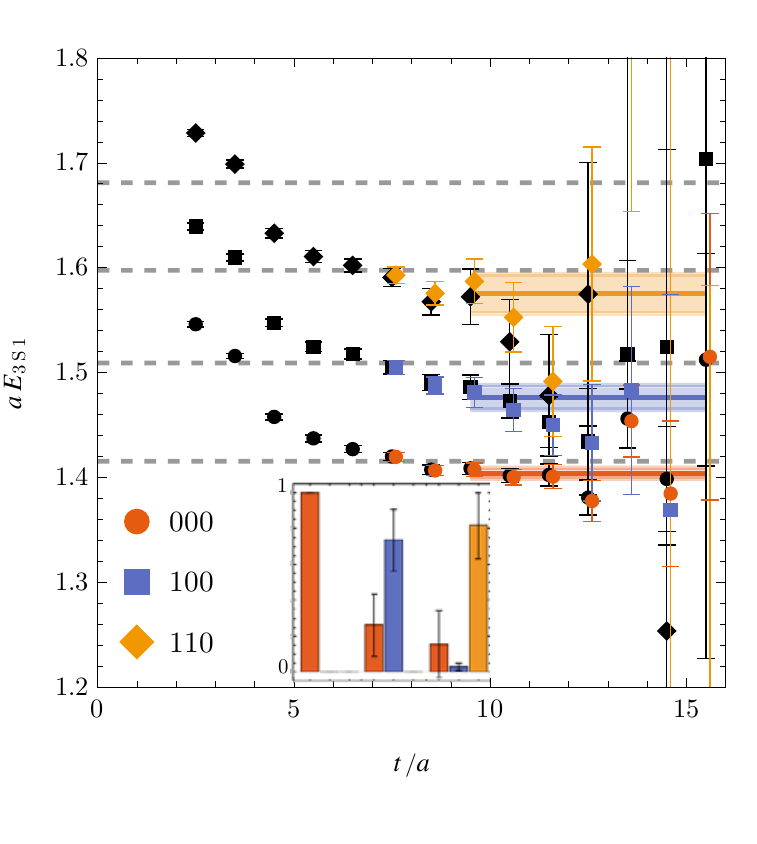}
        \caption{Effective mass curves (upper panels) and coupling factors (lower panels) from the variational analysis for the $^1\mathrm{S}_0$ and $^3\mathrm{S}_1$ channels. Black points denote the raw data, while colored points indicate results after applying the variational method. The bar charts display the overlap factors
$\abs{Z^i_n}^2\equiv \abs{\Braket{\Omega|O_i(t)|n}}^2$, as defined in Eq.~(\ref{eq:overlapFactor}).}
        \label{pic:2ptGEVP}
    \end{figure}

    As discussed in \cref{subsec:interpolators}, dibaryon interpolators can be further optimized using variational methods. In our variational analysis, we include three lowest momentum states. Fig.~\ref{pic:2ptGEVP} compares effective mass curves before and after applying the variational method.
 The overlap factors $Z_{ni}^2\equiv\abs{\Braket{\Omega|O_i|n}}^2$ shown in Fig.~\ref{pic:2ptGEVP}
 indicate that the GEVP ground state couples overwhelmingly ($>99\%$) to zero-momentum operators. This observation suggests that zero-momentum bi-local interpolators are sufficient to approximate the dinucleon ground state for matrix element calculations. However, the situation is different for the extraction of scattering phase shifts.
 The energy difference between the GEVP ground state and the result directly from zero-momentum interpolator is at the percent level of the nucleon mass—a non-negligible correction, given that finite-volume energy shifts appear at comparable magnitudes.

     \begin{figure}[htbp]
        \centering
        \includegraphics[width = 0.89\linewidth]{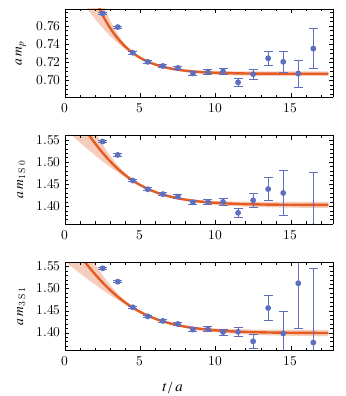}
        \includegraphics[width = 0.89\linewidth]{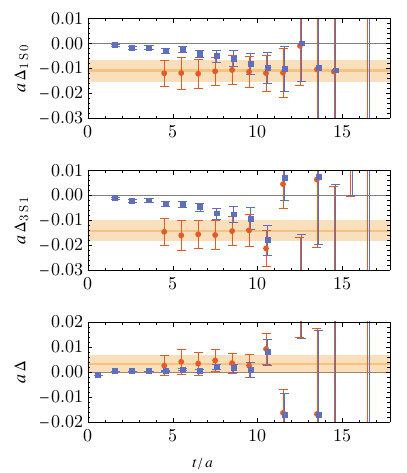}
        \caption{Upper panels: Two-state fits to the effective masses of the proton, diproton ($^1\mathrm{S}_0$), and deuteron ($^3\mathrm{S}_1$).
Lower panels: Energy shifts extracted using the ratio method (blue), subtracted correlators (orange), and the variational method (yellow band).}
        \label{pic:2ptMultiFit}
    \end{figure}

    The variational analysis is performed using dibaryon operators with multiple momenta. Residual contamination from single-nucleon excitations may persist even after applying the variational method. To address this possibility, we perform a two-state fit of the form
    \begin{equation}
        C(t) = A_0 \ce^{-m_0 t}\left( 1 + r \ce^{-\delta t}\right),
        \label{eq:multiStateFit}
    \end{equation}
    where $A_0 \ce^{-m_0 t}$ represents the ground-state contribution, and $r \ce^{-\delta t}$ effectively parameterizes remaining excited-state contamination.
    The upper panel of Fig.~\ref{pic:2ptMultiFit} shows the quality of multi-state fit, while the lower panel compares three analysis methods. The ratio method, $R_{^1\mathrm{S}_0}(t)\equiv C_{^1\mathrm{S}_0}(t)/C_p^2(t)$, shown as blue points, exhibits pronounced time dependence. This behavior is particularly concerning because nuclear energy shifts ($\Delta\sim\mathcal{O}(1\%)M$ with $M$ the nucleon mass) are comparable in magnitude to potential systematic shifts arising from residual excited-state contamination. Notably, this time dependence persists even when standard single-state plateau appear stable, highlighting the danger of pseudo plateau in precision nuclear calculations.
    Having performed a two-state fit, we construct excited-state-subtracted correlators
    \begin{equation}
        \tilde{C}(t) = C(t) - A_0 r \ce^{-(m_0+\delta) t},
    \end{equation}
    where $A_0, r, m_0, \delta$ are determined from the two-state fit. The resulting ratios, shown as orange points in Fig.~\ref{pic:2ptMultiFit}, exhibit markedly improved plateau behavior once residual contamination is removed. This cross-validation confirms that the systematic shifts observed in the blue data originate from uncompensated excited-state contributions.
A comparison between the excited-state-subtracted ratios and the variational results (yellow bands) shows good agreement at large Euclidean times. This consistency allows us to determine a conservative fitting window,
 $t_i=9\,a\approx 1.00$ fm and $t_f=16\,a\approx1.8$ fm, which balances statistical uncertainties against systematic contamination.

 Our final results demonstrate that:
(i) multi-state and variational analyses yield consistent energy shifts $\Delta_{^1\mathrm{S}_0}$ and $\Delta_{^3\mathrm{S}_1}$, differing by less than $<0.5\%$ compared to the nucleon mass; (ii) direct ratio methods suffer from systematic biases when early time slices are included;
(iii) the key $^1\mathrm{S}_0$-$^3\mathrm{S}_1$ splitting $\Delta_{^1\mathrm{S}_0, ^3\mathrm{S}_1}=5.9(6.4)$ MeV is consistent within $1\sigma$ across all methods, with the $^3\mathrm{S}_1$ state having the lower energy.

    We present the results obtained using the variational method with $t_i/a=9$ and $t_f/a=16$ as our final results:
    \begin{align}
        &m_p = \SI{1.2649(42)}{GeV}, \\
        &m_{^1\mathrm{S}_0} = \SI{2.5104(90)}{GeV}, \\
        &m_{^3\mathrm{S}_1} = \SI{2.5045(89)}{GeV}, \\
        &\Delta_{^1\mathrm{S}_0} \equiv m_{^1\mathrm{S}_0} - 2m_p = -19.4(7.9)\,\mathrm{MeV}, \\
        &\Delta_{^3\mathrm{S}_1} \equiv m_{^3\mathrm{S}_1} - 2m_p = -25.3(7.6)\,\mathrm{MeV}, \\
        &\Delta_{^1\mathrm{S}_0, ^3\mathrm{S}_1} \equiv m_{^1\mathrm{S}_0} - m_{^3\mathrm{S}_1} = 5.9(6.4)\,\mathrm{MeV}.
    \end{align}

    \begin{figure}[htbp]
            \centering
            \includegraphics[width = \linewidth]{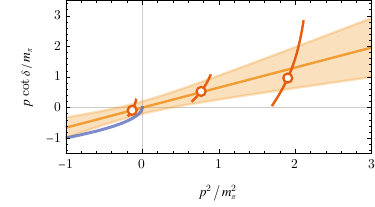}
            \includegraphics[width = \linewidth]{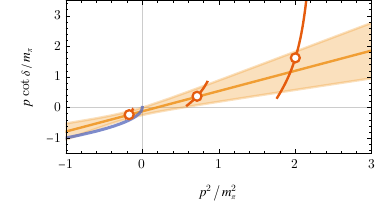}
            \caption{Scattering phase shift analysis for $^1\mathrm{S}_0$ (upper) and $^3\mathrm{S}_1$ (lower) channels.}\label{pic:ERE}
    \end{figure}

    Using ERE~(\ref{eq:ERE}), we determine the scattering length $a$ and effective range $r_0$ using the lowest three GEVP states:
    \begin{align}
        [a^{(pp)}]^{-1} &= \SI{0.02(44)}{fm^{-1}},\quad r_0^{(pp)} = \SI{0.60(28)}{fm}, \label{eq:ERE_pp}\\
        [a^{(d)}]^{-1} &= \SI{0.28(27)}{fm^{-1}},\quad r_0^{(d)} = \SI{0.60(26)}{fm}. \label{eq:ERE_d}
    \end{align}
    We note that the excited states beyond the ground state ($n=1,2$) lie above the $t$-channel cut at $p^2 = (m_\pi/2)^2$, where \textit{inelastic scattering contributions} arising from pion exchange may affect the phase-shift analysis. Such effects are not included in the present calculation.

    For the diproton/dineutron ($^1\mathrm{S}_0$) channel, a physical bound state would correspond to the ERE curve intersecting the bound-state condition $p\cot\delta=-\sqrt{-p^2}$ (blue curve) in the $p^2<0$ region.
    Within current uncertainties, the results are consistent with both a shallow bound state and an attractive scattering state.
In contrast, the deuteron ($^3\mathrm{S}_1$) channel exhibits stronger binding, with central values favoring a shallow bound state, although an attractive scattering state cannot be excluded given the uncertainties.

\subsection{Proton-Proton Fusion}
    In this subsection, we first compute the single-nucleon three-point function to determine the nucleon axial charge $ g_A $.
    The calculation is relatively straightforward and has already been carried out at the physical point with sub-percent precision~\cite{chang2018per}. A comprehensive summary is provided in the Flavour Lattice Averaging Group (FLAG) review~\cite{FlavourLatticeAveragingGroupFLAG:2024oxs}, together with comparisons to the experimentally recommended values from the PDG.
    Because $g_A$ deviates from its physical value at unphysical quark masses, lattice calculations must determine $g_A$ at those masses in order to quantify the ratio of two-body to one-body contributions in two-nucleon systems. This ratio provides a crucial input for determining the two-body LEC $L_{1,A}$.

\begin{figure}[htbp]
        \centering
        \includegraphics[width = \linewidth]{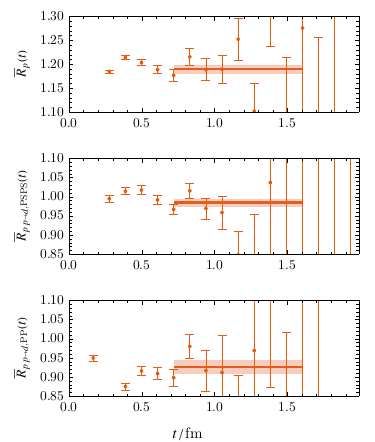}
        \caption{Results for $ g_A $ (top) and the proton-proton fusion matrix elements (middle and bottom panels).
        Here we define $ \overline{R}_3(t) \equiv R_3(t+0.5)-R_3(t-0.5) $.
        The three-point correlation functions for the dibaryon system are normalized by $ g_A $.
        PSPS denotes the use of bi-local interpolators for both the initial and final states, while PP indicates the use of hexa-quark interpolators in both states.
        }
        \label{pic:threePointsResults}
    \end{figure}

Using the relation in Eq.~(\ref{eq:ratioThreePoint}), the weak matrix element of a single proton is extracted from the ratio of three-point to two-point correlation functions, shown in the upper panel of Fig.~\ref{pic:threePointsResults}. The fit yields
    \begin{align}
        g_A
        = 1.190(10),
        \label{eq:gAResult}
    \end{align}
    which is consistent with the value $ g_A = \SI{1.186}(36) $ obtained by the RBC/UKQCD Collaboration on the same lattice ensemble~\cite{Yamazaki_2008}.

    Applying the same ratio method in Eq.~(\ref{eq:ratioThreePoint}) to proton-proton fusion, the corresponding matrix elements are extracted from the ratios of three-point to two-point correlation functions, as shown in the middle and lower panels of Fig.~\ref{pic:threePointsResults}. The resulting fits give
    \begin{align}
        \frac{\abs{\braket{d|\mathcal{J}_3|pp}}_{\mathrm{PSPS}}}{g_A} &= 0.984(10), \label{eq:3ptNME} \\
        \frac{\abs{\braket{d|\mathcal{J}_3|pp}}_{\mathrm{PP}}}{g_A} &= 0.926(18). \label{eq:3ptNMEPP}
    \end{align}
In the absence of rescattering effects and two-body interactions, these ratios are expected to be exactly unity. Deviations from unity therefore reflect contributions from nucleon-nucleon interactions in the proton-proton fusion, providing constraints on the LEC $L_{1,A}$, which governs the leading-order two-body axial current in $\cancel{\pi}\mathrm{EFT}$~\cite{Savage2017Proton-Proton,Davoudi:2021noh}.

A $3.4\sigma$ discrepancy is observed between the two interpolator choices, which is smaller than that seen in the energy spectrum. This reduction arises because excited-state contamination partially cancels in the ratio of three-point to two-point functions. Nevertheless, residual excited-state contamination in both the initial and final states remains non-negligible and can still affect the reliability of the extracted matrix elements, , highlighting the importance of the interpolator choice.
We therefore adopt the result obtained with the bi-local interpolator, which exhibits better control over excited-state contamination, as our preferred estimate of the matrix element.
Following the notation of Ref.~\cite{Savage2017Proton-Proton}, the short-distance two-body contribution (prior to applying the Lellouch-L\"uscher finite-volume corrections) is given by
    \begin{align}
        \frac{\abs{\braket{d|\mathcal{J}_3|pp}}}{g_A} - 1 = -0.016(10). \label{eq:L1Asd-2bData}
    \end{align}
    Our result is consistent within uncertainties but slightly larger in magnitude than the NPLQCD collaboration’s value of $-0.010(01)(13)$ reported in Ref.~\cite{Savage2017Proton-Proton}, where Lellouch-L\"uscher finite-volume corrections were also not included. This level of agreement may be partly accidental, as the two calculations employ different interpolating operators and are performed at different pion masses.

    When Lellouch-L\"uscher finite-volume effects are taken into account, the size of the finite-volume corrections can be estimated using the two-nucleon energy spectrum results discussed in \cref{subsec:spectrum}.
The leading-order (LO) and NLO corrections are summarized in Table~\ref{tab:FVCalcultaion}. The LO result includes only the contribution from the scattering length, while the NLO result additionally incorporates effects from the effective range. The second column of Table~\ref{tab:FVCalcultaion} presents the product of the Lellouch-Lüscher factors for the initial and final states. As discussed in \cref{subsec:FVAnalysis}, these factors can be sizable for two-nucleon systems, a conclusion that is confirmed by our numerical results.
The third column shows the finite-volume one-body contribution calculated using Eq.~(\ref{eq:FV1B}).

    \begin{table}[htbp]
        \centering
        \begin{tabular}{c|cc}
            \hline
            Order of ERE & $L^3\sqrt{\mathcal{R}^{(d)}\mathcal{R}^{(pp)}}$ & $L^3 \Braket{E_f, L|J|E_i, L}_{\mathrm{1B}} / g_A $ \\
            \hline
            LO & $0.52(26)$ & $0.995(11)$ \\
            LO+NLO & $0.64(40)$ & $1.23(19)$ \\
            \hline
        \end{tabular}
        \caption{Lellouch L\"uscher finite-volume factor based on LO and NLO ERE and together with the finite-volume one-body contribution from Eq.~(\ref{eq:FV1B}).}
        \label{tab:FVCalcultaion}
    \end{table}

    A key point is that the directly calculated proton-proton fusion matrix element normalized by $g_A$, $0.984(10)$, although very close to unity, does not imply negligible finite-volume effects.
    Finite-volume corrections affect the determination of the LEC $\tilde{L}_{1,A}$ in two ways. First, the one-body matrix element $\Braket{E_f, L|J|E_i, L}_{\mathrm{1B}}$ must be subtracted to isolate the two-body contribution, and this quantity is itself determined using information from dibaryon scattering. Second, as shown in Eq.~(\ref{eq:FV2B}), Lellouch-L\"uscher factors must be applied to convert
    the two-body matrix element from finite volume to infinite volume. These factors can significantly enhance the infinite-volume matrix element - potentially by a factor of two - although the associated uncertainties remain sizable.

     \begin{figure}[htbp]
        \centering
        \includegraphics[width = \linewidth]{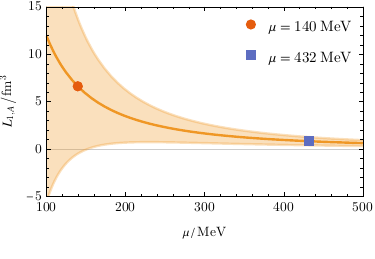}
        \caption{The $\mu$ dependence of $L_{1,A}$. We evaluate $L_{1,A}$ at different renormalization scale $\mu$ using Eq.~(\ref{eq:L1ARelation}) and the input from lattice calculations, with the error propagated from the lattice results. The plot explicitly marks both the points at physical and unphysical pion mass.}
        \label{pic:l1AmuDependence}
    \end{figure}

    Using Eqs.~(\ref{eq:FV1B}), (\ref{eq:FV2B}) and (\ref{eq:W2B}), we extract the scale-independent constant:
    \begin{equation}
        \tilde{L}_{1,A} = \SI{-0.11(12)}{fm^{-1}} = - \SI{22(23)}{MeV}.
    \end{equation}
    The scale-dependent LEC $L_{1,A}$ is then obtained using the relation in Eq.~(\ref{eq:L1ARelation}) together with lattice-QCD inputs. In $\cancel{\pi}\mathrm{EFT}$, a conventional choice of renormalization scale is $\mu = m_\pi$.
    For lattice calculations performed at unphysical quark masses, it is therefore natural to retain this prescription and use the corresponding unphysical pion mass as the scale.
    We observe, however, that $L_{1,A}$ exhibits a strong dependence on $\mu$, approximately scaling as $1/\mu^2$. This behavior originates from the $1/C^{(d)}_0 C^{(pp)}_0$ term and implies that LECs determined at unphysical pion masses cannot be directly compared with experimental extractions or calculations performed at the physical point. To quantify this effect, we evaluate $L_{1,A}$ over a range of renormalization scales $\mu$, as shown in Fig.~\ref{pic:l1AmuDependence}. As $\mu$ decreases, the uncertainty in $L_{1,A}$ increases rapidly, driven by the $(1/a - \mu)$ factors appearing in the denominators of $C_0$ and $C_2$. Reducing this uncertainty will require more precise determinations of two-nucleon scattering parameters. Using the present lattice inputs, we obtain the following constraint on $L_{1,A}$ evaluated at $\mu=m_\pi\big|_{\text{phys}}$
    \begin{align}
        L_{1,A} &= 6.0(7.1)\,\mathrm{fm^3}. \label{eq:L1ALattice}
    \end{align}
    As summarized in Table~\ref{tab:L1AExpData}, existing phenomenological extractions of $L_{1,A}$ from experimental data are subject to sizable uncertainties. In this context, our lattice determination, while still affected by significant uncertainties, is consistent in magnitude with previous results and provides an independent first-principles constraint.

    \begin{table}[htbp]
        \centering
        \begin{tabular}{c|ccc}
            \hline
            Process/Method & $ L_{1,A}/\mathrm{fm}^3 $ \\
            \hline
            SNO \& Super-K~\cite{Chen:2002pv} & $4.0(6.3)$ \\
            reactor $\overline{\nu} d $~\cite{Butler:2002cw} & $3.6(5.5)$ \\
            $\overline{\nu} d $ in $\mathrm{\chi EFT}$~\cite{Acharya:2019fij} & $4.9^{+1.9}_{-1.5}$ \\
            \multirow{2}{*}{Helioseismology~\cite{Brown:2002ih,Butler:2002cw}} & $4.8(6.7)(\mathrm{N^2LO})$ \\
            & $7.0(5.9) (\mathrm{N^4LO})$ \\
            \multirow{2}{*}{Tritium $\beta$ decay~\cite{Schiavilla:1998je,Butler:2002cw}} & $4.2(3.7)(\mathrm{N^2LO})$ \\
            & $6.5(2.4) (\mathrm{N^4LO})$ \\
            Tritium $\beta$ decay~\cite{Nguyen:2024rlr} & $6.01(2.08)$ \\
            Lattice-NPLQCD~\cite{Savage2017Proton-Proton} & $3.9(0.2)(1.0)(0.4)(0.9)$ \\
            \hline
        \end{tabular}
        \caption{Previously reported values of  $L_{1,A} $.}
        \label{tab:L1AExpData}
    \end{table}

 The result in Eq.~(\ref{eq:L1ALattice}) is subject to two dominant sources of uncertainty, in addition to unphysical-mass effects.
First, lattice-QCD uncertainties include both statistical errors and systematic uncertainties associated with the determination of ERE parameters from two-nucleon spectroscopy. In particular, contamination from inelastic scattering above the $t$-channel cut may affect the ERE analysis.
Second, uncertainties arise from the EFT truncation. In the present calculation, the NLO scattering amplitude yields
$\mathcal{M}^{\mathrm{NLO}}/\mathcal{M}^{\mathrm{LO}} \sim 20\% $, indicating a non-negligible higher-order contribution.
These limitations highlight the need for improved control of two-nucleon spectroscopy and EFT convergence.
A further systematic uncertainty arises from finite lattice spacing.
The present calculation is performed at a single lattice spacing $a \simeq 0.11$ fm.
The chiral property of domain-wall fermions naturally reduces the lattice
artifacts to $O(a^2)$; nevertheless, a dedicated continuum-limit study of
the nucleon-nucleon system is important to control this systematic effect.
In Ref.~\cite{Green:2021qol} a study of the $H$ dibaryon at the SU(3)-flavor-symmetric
point found sizable lattice artifacts. We therefore treat discretization effects as an
unquantified systematic uncertainty to be addressed in future calculations at
multiple lattice spacings.

Despite the large uncertainty, this calculation highlights the intrinsic challenges in determining $L_{1,A}$. In particular, the extracted value is highly sensitive to the input two-nucleon scattering length and effective range.
On the lattice, these quantities must be inferred from finite-volume energy shifts, whose determination is notoriously difficult and typically suffers from sizable statistical and systematic uncertainties.
This sensitivity underscores the importance of improving lattice calculations of the two-nucleon scattering system.
More precise determinations of scattering parameters—through larger volumes, improved operator constructions, or moving-frame analyses—would have a substantial impact on reducing the uncertainty of weak two-body matrix elements.

An alternative and complementary strategy is to perform calculations directly at the physical pion mass.
Although such simulations face a deteriorating signal-to-noise ratio, they offer the advantage that experimentally measured scattering lengths and effective ranges can be used as inputs.
This would largely eliminate one of the dominant sources of uncertainty and provide a more robust determination of $L_{1,A}$.

    \section{Conclusion}

    In this work, we employed bi-local interpolators for both initial and final states to compute the two-nucleon energy spectrum and proton–proton fusion matrix elements. We find that bi-local interpolators significantly suppress excited-state contamination, while momentum-space variational analyses demonstrate the dominance of zero-momentum modes in the ground state.

For the energy spectrum, finite-volume energy shifts are below the percent level of the nucleon mass, implying that even small admixtures of higher-momentum modes can noticeably affect the results. In matrix-element calculations, operator optimization using the GEVP ground-state eigenvector ensures that residual excited-state contributions are parametrically suppressed by their overlap factors. Moreover, since matrix elements are extracted from ratios of three-point to two-point functions, excited-state effects partially cancel, leading to smaller interpolator dependence than observed in the energy spectrum.
Nevertheless, the small energy gap between the ground state and rescattering-induced excited states necessitates long Euclidean time separations to achieve reliable plateaus. Operator optimization is therefore essential for controlling systematic uncertainties, a requirement that becomes increasingly important at lighter pion masses due to worsening signal-to-noise ratios.

Using the finite-volume formalism for $2+\mathcal{J}\rightarrow 2$ processes~\cite{Briceno:2015tza,Baroni_2019,Davoudi:2020xdv,Davoudi:2020ngi}, we quantified finite-volume corrections to proton-proton fusion matrix elements and extracted the low-energy constant
$L_{1,A}$ at an unphysical pion mass of $m_{\pi} = \SI{432}{MeV}$. Although two-body contributions are parametrically suppressed relative to one-body terms, finite-volume effects on the latter are substantial. Our results show that finite-volume corrections can significantly enhance the extracted two-body contribution.
The methodology developed here provides a systematic framework for controlling excited-state contamination and finite-volume effects in two-nucleon weak processes, and can be directly extended to other reactions such as double-beta decay.

\begin{acknowledgements}

X.F., B.H.J., C.L. and Z.Y.W. were supported in part by NSFC of China under Grants No. 12125501, No. 12550007, No. 12293060 and No. 12293063.
L.J. acknowledges the support of DOE Office of Science Early Career Award DE-SC0021147, DOE grant DE-SC0010339 and DE-SC0026314.
The research reported in this work was carried out using the computing facilities at Chinese National Supercomputer Center in Tianjin.
It also made use of computing and long-term storage facilities of the USQCD Collaboration, which are funded by the Office of Science of the U.S. Department of Energy.

\end{acknowledgements}

\bibliography{dibaryon.bib}
\end{document}